\documentclass[twocolumn,showpacs,amsmath,amssymb]{revtex4}
\input{epsf}

\usepackage{graphicx}
\usepackage{dcolumn}
\usepackage{array}
\usepackage{bigstrut}
\usepackage{longtable}
\usepackage{rotating,booktabs}
\usepackage{booktabs,threeparttable}
\usepackage{color}
\usepackage{bm}
\def\bdot{\mbox{\boldmath{$\cdot$}}}
\usepackage{color}

\begin{document}

\title{Theory for the Rydberg states of helium: Comparison with experiment for the $1s24p\;^1P_1$ state ($n=24$)}

\author{Aaron T. Bondy$^1$}
\author{G. W. F. Drake$^{1,*}$}
\author{Cody McLeod$^1$}
\author{Evan M. R. Petrimoulx$^1$}
\author{Xiao-Qiu Qi$^{2,3}$}
\author{Zhen-Xiang Zhong$^{4,3}$}

\affiliation { $^1$ Department of Physics, University of Windsor, Windsor, Ontario, Canada N9B 3P4}
\affiliation {$^2$ Department of Physics, Zhejiang Sci-Tech University, Hangzhou 310018, China}
\affiliation {$^3$ State Key Laboratory of Magnetic Resonance and Atomic and Molecular Physics, Wuhan Institute of Physics and Mathematics, Innovation Academy for Precision Measurement Science and Technology, Chinese Academy of Sciences, Wuhan 430071, China}
\affiliation {$^4$ Center for Theoretical Physics, School of Physics and Optoelectronic Engineering, Hainan University, Haikou 570228, China}

\begin{abstract}
Recent measurements of the ionization energies of the Rydberg $^1P$ states of helium for principal quantum number $n = 24$ and higher present a new challenge to theoretical atomic physics.
A long-standing obstacle to high precision atomic theory for three-body systems is a rapid loss of accuracy for variational calculations with increasing principal quantum number $n$.  We show that this problem can be overcome with the use of a ``triple" basis set in Hylleraas coordinates.  Nonrelativistic energies accurate to 23 significant figures are obtained with basis sets of relatively modest size (6744 terms).  Relativistic and quantum electrodynamic effects are calculated, including an estimate of terms of order $m\alpha^6$ from a $1/n^3$ extrapolation, resulting in an estimated accuracy of $\pm$1 kHz.
The calculated ionization energy of 5704\,980.348(1) MHz is in excellent agreement with the experimental value 5704\,980.312(95) MHz.  These results establish the ionization energy of the $1s24p\;^1P_1$ state as an absolute point of reference for transitions to lower-lying states, and they confirm an $11\sigma$ disagreement between theory and experiment in the triplet spectrum of helium.
{Results are also given for the $1s24p\;^3P_J$ states in agreement with a recent experiment on the
triplet Rydberg series, thereby confirming a discrepancy of $0.468\pm0.055$ MHz for the ionization energy of the $1s2s\;^3S_1$ state.}
\end{abstract}

\date{\today}

\pacs{31.15.ap, 31.15.ac, 32.10.Dk} \maketitle

Recent measurements for the ionization energies of the Rydberg $^1P$ states of helium for principal quantum number $n = 24$ to 100 \cite{Clausen} present a new challenge to high-precision atomic theory.
Results for the nonrelativistic wave functions and energies are readily obtained for the low-lying sates of helium and other three-body systems by means of variational calculations in Hylleraas coordinates.  Relativistic and quantum electrodynamic corrections can then be included by perturbation theory \cite{Yan1995,Pachucki2012,Pachucki2010,Zheng2017,Kato2018}.  A long-standing obstacle to further theoretical progress is that the accuracy typically declines quickly with increasing $n$. The problem has now become urgent with the publication of the Clausen et al.\ results \cite{Clausen}.  Their measurements provide absolute point of reference for transitions to lower-lying states where there is an $11\sigma$ disagreement between theory and experiment \cite{Clausen,Patkos2021},  but there is no corresponding theory for comparison.

 The rapid decline in accuracy is already evident in the original ground-breaking calculations of Accad, Pekeris and Schiff \cite{Accad71}.
 Their result for the $1s5p\;^1P$ barely exceeds the accuracy of a simple screened hydrogenic approximation $-2.02000$ E$_{\rm h}$.  Drake et al.\ \cite{Drake87,Drake88,DrakeMakowski88,DrakeYan92} showed that much improved accuracy could be obtained by ``doubling" the basis set so that it explicitly contains two variationally determined distance scales for the inner and outer electron.

Double basis sets proved adequate to obtain results of useful accuracy (parts in $10^{16}$) for all states of helium up to $n = 10$ and angular momentum $L = 7$, including relativistic and QED corrections.  The only calculations to beyond $n = 10$ are the ICI calculations of Nakashima et al.\ \cite{Nakashima2008} up to $n= 24$ for the nonelativistic energies of $S$-states.  In addition, very high precision calculations have been performed by Aznabaev et al.\ \cite{Aznabaev2018} for the nonrelativistic energies for all states up to $n = 4$, using stochastic all-exponential basis functions with up to 22000 terms of the form $\exp(-\alpha_ir_1-\beta_ir_2-\gamma_ir_{12})$.

If doubling the basis set is good, then perhaps tripling it is even better.  Preliminary experiments for the ground states of Ps$^-$, H$^-$ and He \cite{Nistor2002} showed that one does indeed obtain results of improved accuracy for basis sets of the same total size, %and a recent critical assessment \cite{Sati2024} showed that results for the ground state of H$^-$ are comparable with the best in the literature \cite{Aznabaev2018},
but with improved numerical stability.  The most accurate results to date for H$^-$ have been obtained by Aznabaev et al.\ \cite{Aznabaev2018}.

 To be specific, the heliumlike Hamiltonian is diagonalized in a basis set constructed from basis functions of the form (for $P$-states)
 \begin{eqnarray}
 \phi_{i,j,k}(\alpha,\beta;{\bf r}_1,{\bf r}_2) &=& r_1^i\,r_2^{j+1}\,r_{12}^k\exp(-\alpha r_1 - \beta r_2)\cos\theta_2\nonumber\\
                                                &&\mbox{}\pm \mbox{ exchange}
 \end{eqnarray}
where ${\bf r}_1$ and ${\bf r}_2$ are the position vectors of electron 1 and electron 2 relative to the nucleus of charge $Z$, $\cos\theta_2$ gives the angular dependence on ${\bf r}_2$ appropriate to a $p$-electron, and $r_{12} = |{\bf r}_1-{\bf r}_2|$ is the inter-electron separation.  The ``exchange" term denotes an interchange of the labels 1 and 2, with the (+) sign for triplets and the ($-$) sign for singlets.  A variational wave function in a double or triple basis set then has the form
\begin{equation}
\psi({\bf r}_1,{\bf r}_2) = \sum_{p=1}^{2{\rm\, or\, }3}\,\,\sum_{i,j,k}^{i+j+k \le \Omega} c_{i,j,k}^{(p)}\phi_{i,j,k}(\alpha_p,\beta_p;{\bf r}_1,{\bf r}_2)
\end{equation}
where the $c_{i,j,k}^{(p)}$ ar linear variational coefficients, and $\Omega$ controls the size of the basis set such that $i+j+k\le \Omega$. The upper limit for the sum over $p$ is 2 for a double basis set and 3 for a triple basis set.  The four or six nonlinear variational parameters $\alpha_p$, $\beta_p$ are determined by minimizing the energy on a four- or six-dimensional energy surface.  This is accomplished by calculating analytically the derivatives $\partial E/\partial \alpha_p$, $\partial E/\partial \beta_p$ \cite{DrakeMakowski88} and finding their zeros by Newton's method.  The optimization produces a natural separation of the nonlinear parameters into short-range, intermediate, and asymptotic long-range sectors \cite{Sati2024}. The basis set also includes the screened hydrogenic term
$\psi_{1s}({\bf r}_1;Z)\psi_{24p}({\bf r}_2;Z-1) \pm$ exchange for effective nuclear charges $Z$ and $Z-1$ respectively.

The rationale for improved accuracy is that a point of diminishing returns (and numerical instability) is reached with increasing $\Omega$, and so doubling or tripling the basis set allows more terms to be added without $\Omega$ becoming excessively large.

The main purpose of the present work is to extend triple basis set calculations to the high-lying $P$-states, including relativistic and QED corrections, and compare with the high-precision measurements of Clausen et al.\ \cite{Clausen}.  The principal computational step  to obtain nonrelativistic wave functions and energies is to solve the generalized eigenvalue problem
\begin{equation}
{\bf H}\Psi = \lambda{\bf O}\Psi
\end{equation}
where {\bf H} is the Hamiltonian matrix with matrix elements $\langle\phi_{i^\prime,j^\prime,k^\prime}|H|\phi_{i,j,k}\rangle$,  {\bf O} is the overlap matrix in the nonorthogonal basis set, and $\Psi$ is the column vector of basis function coefficients $c_{i,j,k}^{(p)}$.  The Hamiltonian in center-of-mass (CM) coordinates is given by (in atomic units with $4\pi \epsilon_0 = 1$)
\begin{equation}
\label{eq:H}
H = \frac{p_1^2 + p_2^2}{2\mu} +\frac{{\bf p}_1\bdot {\bf p}_2}{M}
-\frac{Ze^2}{r_1} - \frac{Ze^2}{r_2} + \frac{e^2}{r_{12}}
\end{equation}
where $M$ is the nuclear mass, $\mu = Mm_e/(M + m_e)$ is the reduced electron mass, and the ${\bf p}_1\bdot{\bf p}_2/M$ term is the mass polarization operator resulting from the motion of the nucleus in the
CM frame. The correction due to the finite nuclear mass is obtained by comparison with the infinite nuclear mass case $M = \infty$.  The power method is used to find the eigenvalue closest to an initial guess.  The strategy is to perform a sequence of calculations with increasing values of $\Omega$ and assess the rate of convergence to determine the accuracy.

The actual basis sets used in the calculations have two important modifications relative to a simple tripling.  First, for the special case of $P$-states, it is advantageous to include a fourth sector with the roles of the $s$- and $p$-electrons interchanged.  This takes into account mixing of these terms by the ${\bf p}_1\bdot{\bf p}_2/M$ mass polarization interaction. Second, since each combination of powers $i,j,k$ is included four times in all, judicious truncations can be introduced to reduce the total size of the basis set. The truncations are (1) for a given overall $\Omega$, the values in the four sectors are $\Omega_1 = \Omega$, $\Omega_2 = \Omega-4$, $\Omega_3 =\Omega-3$, and $\Omega_4 = \Omega-11$; (2) in sectors 1 and 2, terms are omitted with high powers of $r_1$ and $r_{12}$ such that $i > 3$ and $j > 3$; and (3) in sectors 3 and 4, terms are omitted for which $i+j+k + |i-j| \ge \Omega_i$ for $k > 2$ \cite{Kono83}. These truncations do not affect the ultimate convergence since all powers are eventually included as $\Omega$ increases.  However, for the largest $\Omega = 34$, it reduces the nominal basis set size of 28\,560 terms to a more manageable 6744 terms. The various parameters for the four sectors are summarized in TABLE I.  These basis sets have excellent numerical stability so that standard quadruple precision (32 decimal digit) is normally sufficient, but for $n$ as high as 24, it is necessary to use extended precision (70 decimal digit) arithmetic \cite{Bailey}.

\begin{table}[tb]
\caption{Parameters for the largest ($\Omega = 34)$ basis set.  $N$
is the number of terms in each sector for a total of 6744 terms.  KH is the Kono-Hattori truncation as defined in the text.}
\begin{ruledtabular}
\begin{tabular}{ccccc}
 \multicolumn{1}{c}{$p = $}
&\multicolumn{1}{c}{1}   &\multicolumn{1}{c}{2} & 3 & 4 \\
\hline
$\alpha_p(\ell=0)$  &  1.0000  & 0.06659  & 1.15814  & 4.38531 \\
$\beta_p(\ell=1)$    &  0.02368 & 0.99677   & 0.45288  & 4.61328\\
$\Omega_p$          &  34      &  30      & 31       &   23    \\
$i_{\rm max}$        &  3       &  3       & 31       &   23   \\
$j_{\rm max}$        &  3       &  3       & 31       &   23   \\
KH Trunc.           &  No      &  No      & Yes      &   Yes   \\
$N$                 &  512     &  448     & 3954     &  1830   \\
\end{tabular}
\end{ruledtabular}
\end{table}

TABLE II shows the convergence pattern for the nonrelativistic energy of the $1s24p\;^1P$ state of helium for the case of infinite nuclear mass.  As an indication of the rate of convergence with increasing $\Omega$, the last column shows the ratios ($R$) of successive differences.  Approximately one additional significant figure is obtained in the energy for every increase of $\Omega$ by 2. The extrapolated value has evidently converged to 22 figures after the decimal with an uncertainty of 6 in the 23rd figure.  The convergence pattern is similar for the case of finite nuclear mass. Assuming the exact value $\mu/M=1.370\,745\,634\,614\times10^{-4}$ for $^4$He \cite{NIST}, the result is (in units of E$_{\rm h}$
 throughout)
\begin{equation}
E(^4{\rm He}) = -2.000\,867\,176\,793\,032\,167\,860\,74(6) \mbox{ E$_{\rm h}$}.
\end{equation}
The convergence pattern is similar for the $1s24p\;^3P$ state with the result for $E(^\infty{\rm He})$
\begin{eqnarray*}
 &&-2.000\,873\,014\,566\,616\,659\,392\,25\quad\mbox{upper bound}\nonumber\\
                    &&-2.000\,873\,014\,566\,616\,659\,392\,40(9)\quad\mbox{extrap.}
\end{eqnarray*}
This degree of accuracy for the energy is much more than what is needed in comparison with experiment, but other quantities typically converge much more slowly with to no more than half as many significant figures.
\begin{table}[tb]
\caption{Convergence table for the nonrelativistic energy of
the $1s24p\;^1P_1$ state assuming infinite nuclear mass.  $R$ is the ratio of successive differences.}
\begin{ruledtabular}
\begin{tabular}{lllc}
\multicolumn{1}{c}{$\Omega$}   &\multicolumn{1}{c}{$N$} & \multicolumn{1}{c}{$E$ (E$_{\rm h}$)} & $R$ \\
\hline
%23 & 2260 &  --2.000\,867\,180\,846\,170\,076\,504\,05   &     \\
%24 & 2544 &  --2.000\,867\,180\,846\,170\,094\,121\,11   &     \\
25 & 2836 &  --2.000\,867\,180\,846\,170\,108\,853\,84   & \\%1.20\\
26 & 3168 &  --2.000\,867\,180\,846\,170\,110\,491\,31   & \\%9.00\\
27 & 3508 &  --2.000\,867\,180\,846\,170\,111\,108\,93   & 2.65\\
28 & 3892 &  --2.000\,867\,180\,846\,170\,111\,253\,81   & 4.26\\
29 & 4284 &  --2.000\,867\,180\,846\,170\,111\,268\,37   & 9.95\\
30 & 4724 &  --2.000\,867\,180\,846\,170\,111\,279\,39   & 1.32\\
31 & 5172 &  --2.000\,867\,180\,846\,170\,111\,280\,83   & 7.68\\
32 & 5672 &  --2.000\,867\,180\,846\,170\,111\,281\,75   & 1.57\\
33 & 6180 &  --2.000\,867\,180\,846\,170\,111\,282\,04   & 3.16\\
34 & 6744 &  --2.000\,867\,180\,846\,170\,111\,282\,19   & 1.91\\
\hbox to 0pt{Extrap.} &     &  --2.000\,867\,180\,846\,170\,111\,282\,23(6)& 3.62\\
\end{tabular}
\end{ruledtabular}
\end{table}

With the nonrelativistic wave functions in hand, it is a straightforward matter to calculate the relativistic and QED corrections as expectation values using the standard methods of nrQED.  Since all these terms decrease approximately in proportion to $1/n^3$, they are strongly suppressed for $n=24$ relative to the low-lying states. To order $\alpha^2$ Ry ($\alpha^4mc^2$), the leading terms are the well-known Breit interaction terms (superscripts denote the power of $\alpha$)
(in atomic units) \cite{DrakeYan92,Bethe}
\begin{equation}
E_{\rm rel}^{(2)} = \sum_{i=1}^5 \langle B_i\rangle + \frac{\mu}{M}(\tilde\Delta_2 + \tilde\Delta_{3Z})
\end{equation}
where $B_1 = -(p_1^4+p_2^4)/(8m^3c^2)$ arises from the relativistic variation of mass with velocity, $B_2$ is the orbit-orbit interaction defined by
\begin{equation}
B_2 = -\frac{e^2}{2(mc)^2}\left[\frac{{\bf p}_1\bdot {\bf p}_2}{r_{12}} + \frac{{\bf r}_{12}\bdot({\bf r}_{12}\bdot{\bf p}_1){\bf p_2}}{r_{12}^3}\right]
\end{equation}
$B_3 = B_{\rm so} + B_{\rm soo}$ contains the spin-orbit and spin-other-orbit interactions defined by
\begin{eqnarray}
B_{\rm so} &=& \frac{Z\mu_{\rm B}e(1+2a_{\rm e})}{mc}\left[\frac{{\bf r}_1\times{\bf p}_1\bdot{\bf s}_1}{r_1^3} + \frac{{\bf r}_2\times{\bf p}_2\bdot{\bf s}_2}{r_2^3}\right]\\
%\frac{{\bf r}_{12}\times {\bf p}_2}{r_{12}^3}\right]\bdot{\bf s}_1\right.\nonumber\\
%  &&\left.\mbox{}+ \left[\frac{Z{\bf l}_2}{r_2^3} + \frac{{\bf r}_{21}\times {\bf %p}_1}{r_{12}^3}\right]\bdot{\bf s}_2\right\}
B_{\rm soo} &=& -\frac{\mu_{\rm B}e}{2mcr_{12}^3}{\bf r}_{12}\times\left[(3+4a_{\rm e}){\bf p}_-\bdot{\bf s}_+ -
{\bf p}_+\bdot{\bf s}_-\right]
\end{eqnarray}
where ${\bf p}_\pm = {\bf p}_1\pm{\bf p_2}$, ${\bf s}_\pm = {\bf s}_1\pm{\bf s_2}$, $\mu_{\rm B} = e\hbar/(2mc)$ is the Bohr magnetron, and $a_{\rm e} \simeq\alpha/(2\pi) -0.328\,429\alpha^2$ is the electron anomalous magnetic moment.
$B_5$ is the spin-spin term defined
%\begin{eqnarray}
%B_5 &=& 4\mu_e^2\left[\frac{8\pi}{3}\delta(r_{12}){\bf s}_1\bdot{\bf s}_2\nonumber\right.\\
%&&\mbox{}\left.+
%\frac{{\bf s}_1\bdot{\bf s}_2}{r_{12}^3} - \frac{3({\bf s}_1\bdot{\bf r}_{12}){\bf s}_2\bdot{\bf %r}_{12}}{r_{12}^5}\right],
\begin{equation}
B_5 = 4\mu_e^2\left[\frac{8\pi}{3}\delta(r_{12}){\bf s}_1\bdot{\bf s}_2+
\frac{{\bf s}_1\bdot{\bf s}_2}{r_{12}^3} - \frac{3({\bf s}_1\bdot{\bf r}_{12}){\bf s}_2\bdot{\bf r}_{12}}{r_{12}^5}\right],
\end{equation}
where $\mu_e = \mu_{\rm B}(1 + a_e)$, and the relativistic recoil terms are
\begin{eqnarray}
\tilde\Delta_2&=&-\frac{Z e^2}{2(mc)^2} \sum_{j=1} ^2 \left[\frac{1}{r_j}{
\bf{p_+}} \bdot {\bf p}_j +\frac{1}{r_j^3}{\bf{r}}_j \bdot ({\bf{r}}
_j \bdot {\bf{p_+}}){\bf p}_j \right],\\
\label{eq:009}
%\end{equation}
%\begin{equation}
\tilde\Delta_{3Z}&=&\frac{2Z\mu_{\rm B}e}{mc}\sum_{i=1} ^2 \frac{1}{r_i^3}{\bf{r}}
_i \times {\bf{p_+}} \bdot {\bf{s}}_i.
\label{eq:010}
\end{eqnarray}
%\begin{equation}
%B_{3e\gamma}=\frac{\alpha^2}{2} \sum_{i\ne j} ^3 \frac{1}{r_{ij}^3} {
%\bf{r}}_{ji} \times {\bf{p}}_i \cdot ({\bf{s}}_i-{\bf{s}}_j) ,
%\label{eq:011}
%\end{equation}
  Finally
 $B_4$ contains the $\delta$-function terms $B_4 = \alpha^2\pi[\frac12\delta(r_1)
+ \frac12\delta(r_2) - \delta(r_{12})]$.  To each of these, there are finite nuclear mass corrections of order $\alpha^2\mu/M$ arising from (i) the mass scaling of each term, and (ii) cross terms between the $B_i$ terms in $E_{\rm rel}^{(2)}$ and the mass polarization operator in Eq.\ (\ref{eq:H}).
The relativistic recoil terms $\tilde\Delta_2$ and $\tilde\Delta_{3Z}$ arise from the transformation of the Breit interaction to CM plus relative coordinates \cite{DrakeYan92,Stone1963}.
In the material to follow, the mass scaling and recoil terms taken together are denoted by $E_{\rm RR,M}^{(2)}$, and the cross terms (ii) by $E_{\rm RR,X}^{(2)}$.

The leading QED corrections of order $\alpha^3$ Ry ($\alpha^5mc^2)$ due to electron self-energy and vacuum polarization can be divided into an electron-nucleus part ($\Delta E_{\rm L,1}$) \cite{Kabir} and an electron-electron part ($\Delta E_{\rm L,2}$) \cite{Araki,Sucher} defined by
\begin{eqnarray}
 E_{\rm L,1}^{(3)} =&&\!\!\!\!\! \textstyle\case43Z\alpha^3[\case{19}{30} -\ln(Z\alpha)^2 - \ln k_0]\langle\delta(r_1) + \delta(r_2)\rangle\\
 E_{\rm L,2}^{(3)} =&&\!\!\!\!\! \textstyle\alpha^3[(\case{89}{15} + \case{14}{3}\ln\alpha - \case{20}{3}{\bf s}_1\bdot{\bf s}_2)\langle\delta(r_{12})\rangle-\case{14}{3}Q]
\end{eqnarray}
where $k_0$ is Bethe's mean excitation energy \cite{Bethe} in units of $Z^2$ Ry.
The advantage of subtracting out the $\ln Z^2$ scaling of the Bethe logarithm and including it instead in the $\ln(Z\alpha)^2$ term is that the value is then close to  the hydrogenic value 2.984\,128\,556 ($\pm$0.5\%) for all states of all light atoms and ions studied \cite{YanDrake2003,PachuckiKomasa2004,Korobov2019,Lesiuk2024}.  Its value for the $1s24p\;^1P$ state of helium can be accurately estimated from the $1/n$ expansions of Drake \cite{Drake2001} or Korobov \cite{Korobov2019} to be 2.984\,128\,212(4).  For $\Delta E_{\rm L,2}$, the $Q$ term
is defined by the improper integral
\begin{equation}
\label{Q0}
Q = \frac{1}{4\pi}\lim_{\epsilon\rightarrow 0}\langle
r_{12}^{-3}(\epsilon)+ 4 \pi (\gamma +\ln \epsilon) \delta({\bf r}_{12})\rangle\,
\end{equation}
where $\epsilon$ is the radius of in infinitesimal sphere that is excluded from the range of integration, and $\gamma$ is Euler's constant.

Contributions of order $\alpha^4$ Ry are more difficult to evaluate.  They have been
calculated in their entirety for the $1s2p \;^1P_1$ state by Pachucki et al.\
\cite{Pachucki2017}, and their result of 8.818 MHz can be used to estimate the
contribution for the $1s24p\;^1P_1$ state.  The dominant contributions come from the
$2\;^1P_1 - 2\;^3P_1$ second-order singlet-triplet mixing term (4.7549 MHz) \cite{Drake-Long},  radiative terms that come from a sum of one- and two-loop contributions denoted by $E_{\rm R1}$ and $E_{\rm R2}$ in Ref.\ \cite{Pachucki2006-P} (1.995 MHz), and an $\alpha^4\ln\alpha$ term (0.2120 MHz) \cite{Drake-Khrip}.  The sum of these terms is 6.962 MHz, leaving a balance $\Delta E^{(4)}$ of $8.818 - 6.962 = 1.856$ MHz that comes from a
mixture of first- and second-order terms that are much more difficult to calculate, as
discussed in detail in Refs.\ \cite{Pachucki2006-P,Pachucki2006-S}.  However, all these terms scale roughly in proportion to $1/n^3$, so one can expect a corresponding contribution of approximately $1.856/12^3 = 1.07$ kHz for the $1s24p\;^1P_1$ state.  This estimate is included in the final results with the entire amount taken as the uncertainty.

\begin{table}[!tb]
\caption{ Contributions to the $1s24p\;^1P_1$ state negative ionization energies of $^4$He}.
\begin{ruledtabular}
\begin{tabular}{l l d}
\multicolumn{1}{c}{Contribution}& Order          &\multicolumn{1}{c}{Value (MHz)}\\
\hline
Nonrelativistic $E_{\rm nr}$               &             &  -5704\,993.752\,9745    \\%
1st.\ order mass pol. $E_{\rm M,1}$        &$\mu/M$      &          26.773\,6188          \\%
2nd.\ order mass pol. $E_{\rm M,2}$        &$(\mu/M)^2$  &          -0.108\,9075\\%
Relativistic $E_{\rm rel}^{(2)}$           &$\alpha^2$   &         -13.292\,1518   \\%
%Anomalous mag.\ mom. (Eanom)    &$\alpha^3$&          0&.000\,0000    \\%
Relativ.\ finite mass $E_{\rm RR,M}^{(2)}$ &$\alpha^2\mu/M$&        -0.006\,7917    \\%
Relativistic recoil $E_{\rm RR,X}^{(2)}$   &$\alpha^2\mu/M$&         0.003\,7235    \\%
e$^{-}$--nucleus QED $E_{\rm L,1}^{(3)}$   &$\alpha^3$   &           0.070\,703    \\%
e$^{-}$--e$^-$ QED $E_{\rm L,2}^{(3)}$     &$\alpha^3$   &          -0.040\,2072    \\%
Singlet-triplet mixing $E_{\rm st}^{(4)}$  &$\alpha^4$   &           0.002\,2762    \\%
Radiative $E_{\rm R1}^{(4)} + E_{\rm R2}^{(4)}$&$\alpha^4$&          0.001\,365  \\ %
Radiative $\alpha^4\ln\alpha$              &$\alpha^4\ln\alpha$ &    0.000\,158     \\
$\Delta E^{(4)}$ scaled from $1s2p\;^1P$   &$\alpha^4$    &          0.001(1)       \\
Finite nuclear size $E_{\rm nuc}$       &$(R/a_0)^2 $     &          0.000\,0433    \\%
Total                                      &              & -5704\,980.348(1)       \\%
Experiment \cite{Clausen}                  &              & -5704\,980.312(95)      \\
Difference                                 &              &         -0.035(95)      \\
\end{tabular}
\end{ruledtabular}
\end{table}
\begin{table}[!tb]
{
\caption{ Contributions to the $1s24p\;^3P_{\rm c}$ (centroid) state negative ionization energies of $^4$He.  The fine-structure splittings are $\Delta E_{2-1} = 1.165(1)$ MHz and $\Delta E_{2-0} = 15.501(4)$ MHz.}
\begin{ruledtabular}
\begin{tabular}{l l d}
\multicolumn{1}{c}{Contribution}& Order          &\multicolumn{1}{c}{Value (MHz)
\footnote{{$E_{\rm centroid} = [E(^3P_0) + 3E(^3P_1) + 5E(^3P_2)]/9$}}}\\
\hline
Nonrelativistic $E_{\rm nr}$               &               &-5743\,372.528\,118 \\%
1st.\ order mass pol. $E_{\rm M,1}$        &$\mu/M$        &       -29.809\,429 \\%
2nd.\ order mass pol. $E_{\rm M,2}$        &$(\mu/M)^2$    &        -0.108\,918 \\%
Relativistic $E_{\rm rel}^{(2)}$           &$\alpha^2$     &         0.962\,174 \\%
Relativ.\ finite mass $E_{\rm RR,M}^{(2)}$ &$\alpha^2\mu/M$&         0.000\,952  \\%
Relativistic recoil $E_{\rm RR,X}^{(2)}$   &$\alpha^2\mu/M$&         0.000\,841  \\%
e$^{-}$--nucleus QED $E_{\rm L,1}^{(3)}$   &$\alpha^3$     &        -0.615\,654  \\%
e$^{-}$--e$^-$ QED $E_{\rm L,2}^{(3)}$     &$\alpha^3$     &        -0.021\,796  \\%
Singlet-triplet mixing $E_{\rm st}^{(4)}$  &$\alpha^4$     &        -0.000\,759  \\%
Radiative $E_{\rm R1}^{(4)} + E_{\rm R2}^{(4)}$&$\alpha^4$ &        -0.010\,857  \\ %
Radiative $\alpha^4\ln\alpha$        &$\alpha^4\ln\alpha$  &         0.0        \\
$\Delta E^{(4)}$ scaled from $1s2p\;^3P_{\rm c}$   &$\alpha^4$     &         0.000\,3(3)  \\
Finite nuclear size $E_{\rm nuc}$       &$(R/a_0)^2 $      &         -0.00\,0418 \\%
Total                                     &                & -5743\,402.131\,7(3) \\%
Experiment \cite{Clausen2025}              &              &  -5743\,402.130(19)      \\
Difference                                 &              &           0.002(19)      \\
\end{tabular}
\end{ruledtabular}
}
\end{table}

The last correction to be included is that due to finite nuclear size.  To lowest order, the energy shift to sufficient accuracy is $E_{\rm nuc} = \frac23\pi Z(R/a_0)^2\langle\delta({\bf r}_1) + \delta({\bf r}_2)\rangle$ where $a_0$ is the Bohr radius and $R=1.6786(12)$ fm \cite{Pachucki2024} is the radius of the nuclear charge distribution for $^4$He.

The various calculated contributions to the ionization energy of the $1s24p\;^1P_1$ state of $^4$He are summarized in TABLE III.  The dominant source of uncertainty ($\pm$1 kHz) is the residual QED contribution of order $\alpha^4$ Ry.  However, this level of theoretical uncertainty is already much better than for any other state of helium, and it establishes an absolute point of reference for transitions to any other state. It is also in excellent agreement with the experimental value 5704\,980.312(95) MHz \cite{Clausen} obtained from a quantum defect fit of ionization energies up to $n = 100$. It is in even better agreement with the value 5704\,980.352(40) MHz obtained directly from the $1s2s\;^1S_0 - 1s24p\;^1P_1$ transition frequency, indicating that the ionization energy of the $1s2s\;^1S_0$ is correctly determined. The corresponding results for the $1s24p\;^3P_{\rm c}$ centroid
are summarized in TABLE IV. Measurements of the $2\;^3S_1 - n\;^3P_{\rm c}$ transition frequencies for the triplet-$P$ states have recently been completed down to $n = 27$ \cite{Clausen2025}.  A quantum defect extrapolation to $n = 24$ is in excellent agreement, as shown in the table, leaving a residual discrepancy of
$\nu_{\rm exp} - \nu_{\rm theo} = 0.468\pm0.055$ MHz for the $2\;^3S_1 - 24\;^3P_{\rm c}$ transition.

The particular significance of this close agreement for the $1s2s\;^1S_0$ state is that it leaves unexplained a $7\sigma$ disagreement of 0.409(61) MHz between theory and experiment for the ionization energy of the $1s2s\;^3S_1$ state \cite{Clausen,Patkos2021}, based on a direct measurement of the $1s2s\;^1S_0 - 1s2s\;^3S_1$ transition frequency \cite{Rooij2011}.  The $11\sigma$ discrepancy of 0.406(36) MHz  is even bigger for the $1s2p\;^3P_1$ state \cite{Clausen}.  {A detailed comparison between theory and experiment is contained in Ref.\ \cite{Clausen}. On the theoretical side, the calculation of Patk\'o\v{s} et al.\ \cite{Patkos2021} greatly reduced the uncertainty for the low-lying $2\;^3S_1$ and $2\;^3P_J$ by a complete calculation of the QED terms of order $\alpha^5$ Ry.  They also estimated the contribution from the next higher-order $\alpha^6$ Ry term, based on a hydrogenic approximation. The result of --0.158(52) MHz for the (positive) ionization energy of the $2\;^3S_1$ state stands as the dominant source of theoretical uncertainty. In contrast, this uncertainty, and the entire $\alpha^4$ Ry term, are suppressed to the 1 kHz level or below by the $1/n^3$ scaling at $n=24$, thereby allowing a clear comparison with experiment for the singlet spectrum ($\pm$95 kHz),  uncomplicated by higher-order QED uncertainties. A parallel experimental study of the triplet-$P$ spectrum \cite{Clausen2025} confirms the discrepancy ($0.468\pm0.055$ MHz).}

In summary, we have extended the ability to perform high precision variational calculations for helium well beyond the previous limit of $n = 10$ \cite{DrakeYan92} to $n = 24$, and performed a complete calculation of relativistic and QED corrections up to order $\alpha^3$ Ry for the $1s24p\;^1P_1$ state,
and estimated the $\alpha^4$ Ry contribution. %An estimate of the $\alpha^4$ terms from the $1/n^3$ scaling from the $1s2p\;^1P_1$ state yields a contribution of only $4\pm 1$ kHz, which is well below the level of experimental accuracy.  Thus the combination of theory and experiment establishes the ionization energy $1s24p\;^1P_1$ state as an absolute point of reference for transitions to lower-lying states.
The success of the present calculation lays the foundation for a comprehensive study of the unexplored territory of Rydberg states between $n = 10$ and 24, and applications to high precision quantum defect extrapolations to higher $n$.  A complete calculation of the terms of order $\alpha^4$ Ry is in progress and will be reported in a future publication.  This, together with other known radiative recoil corrections, will reduce the theoretical uncertainty to well below 1 kHz, where long-range Casimir-Polder effects may become significant \cite{Levin1993}.

\begin{acknowledgements}
This research was supported by the Natural Sciences and Engineering Research Council of Canada (NSERC) and by the Digital Research Alliance of Canada/Compute Ontario.  In addition, X.-Q. Q.\ and Z.-X. Z.\ acknowledge support by the National Natural Science Foundation of China under Grant Nos. 12204412, 11974382, and 12393821, and by the Science Foundation of Zhejiang Sci-Tech University under Grant No. 21062349-Y."
\end{acknowledgements}

$^*$Email address: gdrake@uwindsor.ca


\begin{references}
%Yan1995,Pachucki2012,Pachucki2010,Zheng2017,Kato2018
\bibitem{Clausen}
G. Clausen, P. Jansen, S. Scheidegger, J. A. Agner, H. Schmutz, and F.
Merkt, Ionization energy of
the metastable $2\;^1S_0$ state of $^4$He from Rydberg-series extrapolation, Phys.\ Rev.\ Lett.\ {\bf 127}, 093001 (2021).
\bibitem{Yan1995}
Z.-C. Yan and G. W. F. Drake, High Precision Calculation of Fine
Structure Splittings in Helium and He-Like Ions, Phys.\ Rev.\ Lett.\
{\bf 74}, 4791 (1995).
\bibitem{Pachucki2012}
K. Pachucki, V. A. Yerokhin, and P. Cancio Pastor, Quantum
electrodynamic calculation of the hyperfine structure of $^3$He, Phys.\
Rev.\ A {\bf 85}, 042517 (2012).
\bibitem{Pachucki2010}
K. Pachucki and V. A. Yerokhin, Fine Structure of Heliumlike Ions and
Determination of the Fine Structure Constant, Phys.\ Rev.\ Lett.\ {\bf
104}, 070403 (2010).
\bibitem{Zheng2017}
X. Zheng, Y. R. Sun, J.-J. Chen, W. Jiang, K. Pachucki, and S.-M. Hu,
Laser Spectroscopy of the Fine-Structure Splitting in the $2\;^3P_J$
Levels of $^4$He, Phys.\ Rev.\ Lett.\ {\bf 118}, 063001 (2017).
\bibitem{Kato2018}
K. Kato, T. D. G. Skinner, and E. A. Hessels, Ultrahigh-Precision
Measurement of the $n=2$ Triplet $P_J$ Fine Structure of Atomic Helium
Using Frequency-Offset Separated Oscillatory Fields, Phys.\ Rev.\
Lett.\ {\bf 121}, 143002 (2018).
\bibitem{Patkos2021}
V. Patk\'o\v{s}, V.A. Yerokhin and K. Pachucki, Complete $m\alpha^7$
Lamb shift of helium triplet states, Phys.\ Rev.\ A {\bf 103}, 042809
(2021).
\bibitem{Accad71}
Y. Accad, C. L. Pekeris, and B. Schiff, $S$ and $P$ States of the
Helium Isoelectronic Sequence up to $Z = 10$, Phys.\ Rev.\ A, {\bf 4},
516 (1971).
\bibitem{Drake87}
G. W. F. Drake, New variational techniques for the 1snd states of
helium, Phys.\ Rev.\ Lett.\ {\bf 59}, 1549 (1987).
\bibitem{Drake88}
G. W. F. Drake, High precision variational calculations for the $1s2\
^1S$ state of H$^-$ and the $1s2\ ^1S$, $1s2s\ ^1S$ and $1s2s\ ^3S$
states of helium, Nucl.\ Instrum.\ Methods Phys.\ Res.\ Sect.\ B, {\bf
31}, 7 (1988).
\bibitem{DrakeMakowski88}
G. W. F. Drake and A. J. Makowski, High Precision eigenvalues for the
$1s2p\ ^1P$ and $^3P$ states of helium, J. Opt.\ Soc.\ Am.\ B, {\bf 5},
2207 (1988).
\bibitem{DrakeYan92}
G. W. F. Drake and Z.-C. Yan, Energies and relativistic corrections for
the Rydberg states of helium: Variational results and asymptotic
analysis, Phys.\ Rev.\ A, {\bf 46}, 2378 (1992).
\bibitem{Nakashima2008}
H. Nakashima, Y. Hijikata, and H. Nakatsuji, Solving the electron and
electron-nuclear Schr\"odinger equations for the excited states of
helium atom with the free iterative-complement-interaction method, J.
Chem.\ Phys.\ {\bf 128}, 154108 (2008).
\bibitem{Aznabaev2018}
D. T. Aznabaev, A. K. Bekbaev, and V. I. Korobov, Nonrelativistic
energy levels of helium atoms, Phys.\ Rev.\ A {\bf 98}, 012510 (2018).
\bibitem{Nistor2002}
G. W. F. Drake, M. M. Cassar and Razvan A. Nistor, Ground-state
energies for helium, H$^-$ and Ps$^-$, Phys.\ Rev.\ A {\bf 65}, 054501
(2002).
\bibitem{Sati2024}
E. M. R. Petrimoulx, A. T. Bondy, E. A.  Ene, L. A. Sati, and G. W. F.
Drake, Ground-state energy of H$^-$: a critical test of triple basis
sets, Can.\ J. Phys.\ DOI10.1139/cjp-2023-0277 (2024).

%\bibitem{Frolov}
%A. M. Frolov, Bound state properties and photodetachment of the negatively
%charged hydrogen ions, Eur.\ Phys.\ J. D {\bf 69}, 132 %(2015).

\bibitem{Kono83}
A. Kono and S. Hattori, Variational calculations for excited states in
He I: Improved estimation of the ionization energy from accurate
energies for the $n\;^3S$, $n\;^1D$, $n\;^3D$ series, Phys.\ Rev.\ A
{\bf 31}, 1199 (1985).
\bibitem{Bailey}
DQFUN: A double-quad precision package with special functions,
Available from https://www.davidhbailey.com/dhbsoftware/READM
E-dqfun.txt [accessed: 2010-09-30].
\bibitem{NIST}
National Institute of Standards and Technology,
https://physics.nist.gov/cuu/Constants/index.html.
https://www.davidhbailey.com/dhbsoftware/README-dqfun.txt

\bibitem{Bethe}
H. A. Bethe and E. E. Salpeter, {\it Quantum Mechanics of One- and
Two-Electron Atoms}, (Springer, New York, 1957), p.\ 181.
\bibitem{Stone1963}
A. P. Stone, Proc. Phys. Soc. {\bf 77}, 786 (1961); {\bf 81}, 868
(1963).
\bibitem{Kabir}
P. K. Kabir and E. E. Salpeter, Radiative Corrections to the
Ground-State Energy of the Helium Atom, Phys.\ Rev.\ {\bf 108}, 1256
(1957).
\bibitem{Araki}
H. Araki, Quantum-Electrodynamical Corrections to Energy-Levels of
Helium, Prog.\ Theo.\ Phys.\ {\bf 17}, 619 (1957).

\bibitem{Sucher}
J. Sucher, Energy Levels of the Two-Electron Atom to Order $\alpha^3$
Ry; Ionization Energy of Helium Phys.\ Rev.\ {\bf 109}, 1010 (1958).
\bibitem{YanDrake2003}
Z.-C. Yan and G. W. F. Drake, Bethe Logarithm and QED Shift for
Lithium, Phys.\ Rev.\ Lett.\ {\bf 91}, 113004 (2003).

\bibitem{PachuckiKomasa2004}
K. Pachucki and J. Komasa, Relativistic and QED Corrections for the
Beryllium Atom, Phys.\ Rev. Lett.\ {\bf 92}, 213001 (2004).
\bibitem{Korobov2019}
V. I. Korobov, Bethe logarithm for the helium atom, Phys.\ Rev.\ A {\bf
100}, 012517 (2019).

\bibitem{Lesiuk2024}
M. Lesiuk and J. Lang, Atomic Bethe logarithm in the mean-field
approximation, Phys.\ Rev.\ A {\bf 108}, 042817 (2023).  The values
obtained for $\ln( k_0/Z^2)$ for atoms heavier than Be are larger than
the hydrogenic value by up to 4.7\%.
\bibitem{Drake2001}
G. W. F. Drake, QED Effects in Helium and Comparisons with High
Precision Experiment, Phys.\ Scr.\ {\bf T95}, 22 (2001).

\bibitem{Pachucki2017}
K. Pachucki, V. Patk\'os, and V. A. Yerokhin, Testing fundamental
interactions on the helium atom, Phys.\ Rev.\ A {\bf 95}, 062510
(2017).
\bibitem{Drake-Long}
 G. W. F. Drake in {\it Long-Range Casimir Forces : Theory and Recent Experiments on Atomic Systems},
 Edited by Frank S. Levin and David A. Micha (Plenum Press, New York, 1993), p.\ 169 and TABLE XII.

\bibitem{Pachucki2006-P}
K. Pachucki, Helium energy levels including $m\alpha^6$ corrections, Phys.\ Rev.\ A {\bf 74}, 062510 (2006).  His value for $E_C$ of $0.201363 \alpha^6m$ (3.757 MHz) includes also mixing with higher $n\;^3P_1$ states with $n > 2$.  This reduces the value from our $E_{\rm st}^{(4)} =4.755$ MHz.
\bibitem{Drake-Khrip}
  G. W. F. Drake, I. B. Khriplovich, A. I. Milstein and A. S. Yelkovsky,
Energy corrections of order $mc^2\alpha^6\ln\alpha $ in helium, Phys.\
Rev.\ A {\bf 48}, R15 (1993).
\bibitem{Pachucki2006-S}
K. Pachucki, $\alpha^4{\cal R}$ corrections to singlet states of
helium, Phys.\ Rev.\ A {\bf 74}, 022512 (2006).
\bibitem{Pachucki2024}
K. Pachucki, V. Lensky, F. Hagelstein, S. S. L. Muli, S. Bacca, and R.
Pohl, Comprehensive theory of the Lamb shift in light muonic atoms
Rev.\ Mod.\ Phys. {\bf 96}, 015001 (2024).

\bibitem{Rooij2011}
R. van Rooij, J. S. Borbely, J. Simonet, M. D. Hoogerland, K. S. E.
Eikema, R. A. Rozendaal, and W. Vassen, Frequency Metrology in Quantum
Degenerate Helium: Direct Measurement of the $2\;^3S_1 \rightarrow
2\;^1S_0$ Transition, Science {\bf 333}, 196 (2011).
\bibitem{Clausen2025}
G. Clausen, K. Gamlin, J. A. Agner, H. Schmutz, and F.
Merkt, Metrology in a two-electron atom: The ionization energy of metastable
triplet helium ($2\;^3S_1$), Phys.\ Rev.\ A in press (2025).
\bibitem{Levin1993}
{\it Long Range Casimir Forces: Theory and Recent Experiments in Atomic
Systems}, Edited by Frank S. Levin and David Micha (Plenum, New York,
1993).
\end{references}
\end{document}